\documentclass{article}
\usepackage{spconf,amsmath,graphicx}
\usepackage{graphicx}
\usepackage{amsmath}
\usepackage{amssymb}
\usepackage{booktabs}
\usepackage{mathrsfs}
\usepackage{algorithm}
\usepackage{algorithmic}
\usepackage{algorithm}
\usepackage{algorithmic}
\usepackage{algorithm}
\usepackage{algorithmic}
\usepackage{amsmath,amssymb}
\usepackage[utf8]{inputenc}
\usepackage{graphicx}
\usepackage{booktabs}
\usepackage{witharrows}
\usepackage{rotating}  
\usepackage{multirow}
\usepackage{pifont}
\usepackage{bbding}
\usepackage{adjustbox}
\usepackage{colortbl}
\usepackage{caption}
\usepackage{xcolor}
\definecolor{bb}{rgb}{0.0, 0.0, 0.5}
\def\etal{\emph{et al.}}
\definecolor{Gray}{gray}{0.9}

\newcommand{\ie}{\textit{i.e.}}
\newcommand{\eg}{\textit{e.g.}}
\newcommand{\bihan}[1]{{\color{black} #1}}

\title{Unsupervised Deep Digital Staining for Microscopic Cell Images \\ via Knowledge Distillation}
\name{\vspace{-0.1in} Ziwang Xu$^{1\,*}$, Lanqing Guo$^{1\,*}$\thanks{* co-first authors contributed equally.}, Shuyan Zhang$^2$, Alex C. Kot$^{1}$ and Bihan Wen$^{1\,\dag}$\thanks{$\dag$ Bihan Wen is the corresponding author.}}
\address{$^1$School of Electrical \& Electronic Engineering, Nanyang Technological University, Singapore\\
$^2$Institute of Bioengineering and Bioimaging, A*STAR, Singapore}

\begin{document}
\ninept
\maketitle
\begin{abstract}
Staining is critical to cell imaging and medical diagnosis, which is expensive, time-consuming, labor-intensive, and causes irreversible changes to cell tissues. Recent advances in deep learning enabled digital staining via supervised model training. However, it is difficult to obtain large-scale stained/unstained cell image pairs in practice, which need to be perfectly aligned with the supervision. In this work, we propose a novel unsupervised deep learning framework for the digital staining of cell images using knowledge distillation and generative adversarial networks (GANs). A teacher model is first trained mainly for the colorization of bright-field images. After that, a student GAN for staining is obtained by knowledge distillation with hybrid non-reference losses. We show that the proposed unsupervised deep staining method can generate stained images with more accurate positions and shapes of the cell targets. Compared with other unsupervised deep generative models for staining, our method achieves much more promising results both qualitatively and quantitatively. 

\end{abstract}
\section{introduction}
\label{sec:intro}

Staining is a key process in cell imaging and medical diagnosis where clinicians could evaluate the morphological and chemical information in a microscopic setting based on enhanced imaging contrast.
Hematoxylin and Eosin (H\&E) is the most commonly used dye staining technique~\cite{feldman2014tissue}.
\bihan{However, the H\&E staining process can be quite time-consuming and expensive in practice}, \ie, taking around 45 minutes and costing about \$2–5 per slice~\cite{rivenson2019phasestain}. 
\bihan{Furthermore, though standard H\&E protocols have been established, the cell staining outcomes usually vary in different histopathology laboratories subjective to the specific staining conditions, which degrade the downstream diagnosis. As staining is an irreversible process, one needs to prepare new cell samples and conduct re-staining from scratch once the results are unsatisfied.}

\bihan{To mitigate these limitations of physical staining, recent works proposed deep learning methods for digital staining for microscopy cell images captured by Quantitative phase imaging (QPI) ~\cite{rivenson2019phasestain}, Autofluorescence imaging ~\cite{zhang2020digital,rivenson2019virtual,rivenson2020emerging} and whole slide imaging (WSI)~\cite{rana2020use}.} Compared to these imaging techniques, dark-field imaging rejects the unscattered background light from the sample, thus preserving the best imaging contrast for transparent and thin samples.
\bihan{It remains unclear if deep digital staining can be applied for dark-field cell image staining with its unique image properties and distribution.
Furthermore, existing deep learning methods for digital staining are all supervised~\cite{rivenson2019phasestain,zhang2020digital,rivenson2019virtual,rivenson2020emerging,rana2020use}, \ie, they all require a large-scale dataset of perfectly aligned staining pairs of cell images for model training. In practice, it is extremely challenging to obtain such a dataset because (1) One can hardly maintain the same orientation of cell samples in different imaging trails, which inevitably introduces rotational and translational deviations in the staining image pairs even for same tissue block; (2) The physical staining process always deforms the tissue constituents, making it almost impossible to pair the bright-field and dark-field images precisely. As a result, the stained cell images using the supervised methods will be significantly degraded due to inaccurate pairing, leading to mistakes in the subsequent medical diagnosis.
}

\begin{figure}[!t] 
\centering 
\vspace{-4mm}
\includegraphics[width=0.48\textwidth]{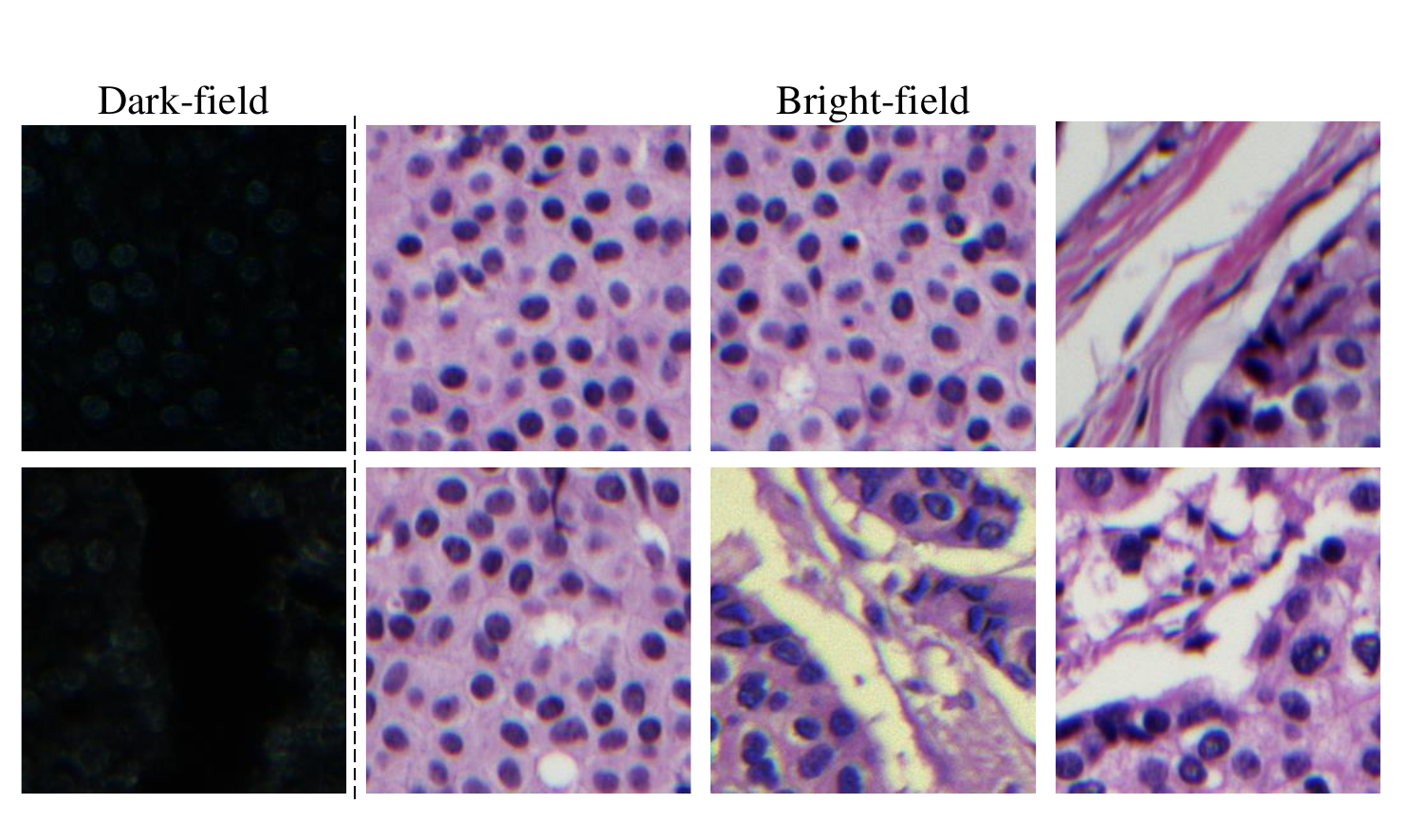} 
\vspace{-5mm}
\caption{Examples of the dark-field and bright-field images from our proposed digital staining dataset (DSD).} 
\label{fig:dataset_samples} 
\vspace{-0.15in}
\end{figure}

\bihan{In this work, we present a novel digital staining scheme for dark-field cell images using \textit{unsupervised deep learning}. Different from all existing digital staining methods, the proposed method does \textbf{not} require any \textit{paired and aligned} unstained/stained images for training. We propose to decompose the staining task into image light enhancement and colorization problems, sequentially. Based on such a staining model, we first enhance the dark-field cell images by matching their illuminance distribution to the target bright-field images, followed by training a teacher model to approximate the light enhancement process. After that, a novel color-inversion distillation is applied to obtain a student generative adversarial network (GAN) with hybrid non-reference losses, which can simultaneously transfer the color style and preserve the consistency of structural information in digital staining.
To our best knowledge, this is the first deep unsupervised learning method for digital staining of dark-field images.}
\begin{figure*}[!t] 
\centering 
\includegraphics[width=1.\textwidth]{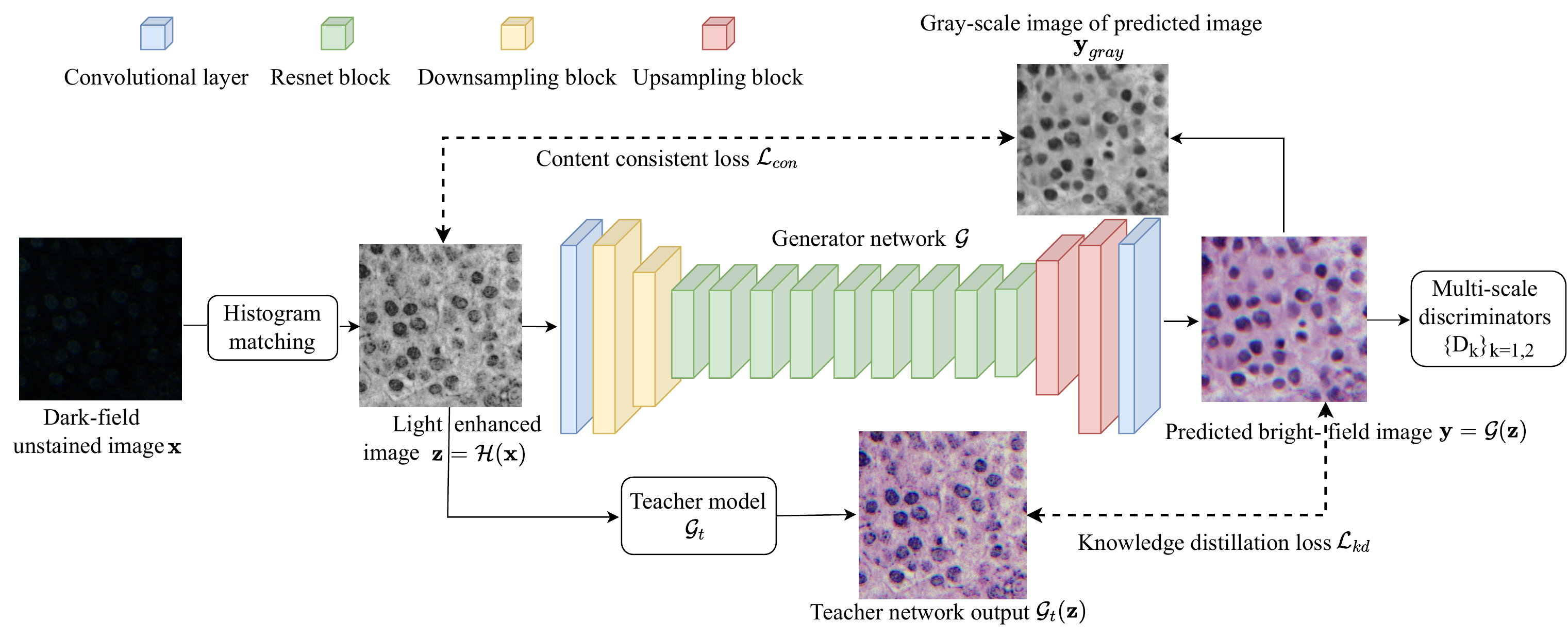}
\vspace{-3mm}
\caption{The overall structure of our method. 
With a dark-field image $\mathbf{x}$, the enhanced image is given by $\mathbf{z}=\mathcal{H}(\mathbf{x})$, then the generator $\mathcal{G}$ outputs the predicted image $\mathbf{y}=\mathcal{G}(\mathbf{z})$. 
Notes that the generator $\mathcal{G}$ also serves as the student network in knowledge distillation.} 
\label{fig:baseline} 
\end{figure*}
\bihan{Furthermore, for benchmarking various digital staining methods for dark-field microscopy cell images, we propose the first real cell-image Digital Staining Dataset (DSD), including the sets of unstained dark-field and stained bright-field images, respectively, without pairing. Fig.~\ref{fig:dataset_samples} shows some examples, with diverse tissue structures, in our dataset.
Experimental results demonstrate the effectiveness of our method with its superior performance over the proposed staining dataset compared to the popular unsupervised deep generative models.
}
Our major contributions can be summarized as follows:
\begin{itemize}
     \item A new dataset for digital cell image staining with unpaired real dark-field unstained and bright-field stained images.
    
    \item \bihan{A novel deep digital staining method for dark-field cell images, which incorporates GAN and knowledge distillation.}
    
    \item \bihan{An effective implementation of deep digital staining which provides state-of-the-art results without supervised training.}

\end{itemize}    

\section{RELATED WORKS}
\label{sec:format}
\noindent\textbf{Digital staining.} 
Learning-based methods achieve superior performance on digital staining recently, which have explored different modalities for unstained images, \eg, quantitative phase microscopy images~\cite{rivenson2019phasestain}, autofluorescence images~\cite{zhang2020digital,rivenson2019virtual,rivenson2020emerging}, and whole slide images~\cite{rana2020use}.
However, these methods are all supervised, i.e., using models that are trained on aligned staining image pairs.

\vspace{1mm}
\noindent\textbf{Image colorization.}
Image colorization aims to learn a mapping function from a grayscale image to the corresponding color counterpart, which is a typical ill-posed inverse problem since one grayscale image may correspond to multiple potential color outputs.  As the development of deep neural networks in recent years, colorization methods are generally learning based~\cite{zhang2016colorful,zhang2017real,su2020instance,wu2021towards}. For instance, the pioneer works~\cite{zhang2016colorful,zhang2017real} proposed an effective backbone with a simple pixel-wise $\ell_1$ loss. After that, Wu~\etal~\cite{wu2021towards} employed generative color prior as guidance. Su~\etal~\cite{su2020instance} used pre-trained networks for object detection to achieve better semantic representation.

\vspace{1mm}
\noindent\textbf{Knowledge distillation for image-to-image translation.}
Knowledge distillation intends to promote the training of student model under the supervision of a teacher model. 
Image-to-image translation have been widely used in the area of image processing~\cite{yu2022towards,guo2022shadowdiffusion,wang2023raw}. 
Several works have been proposed about using knowledge distillation in image-to-image translation tasks~\cite{li2020gan,chen2020distilling,jin2021teachers}. Li~\etal~\cite{li2020gan} minimize the Euclidean distance between the latent features from teacher and student models. Chen~\etal~\cite{chen2020distilling} distills the generator and the discriminator simultaneously in GAN. Jin~\etal~\cite{jin2021teachers} conducts distillation on intermediate features of the generator with global kernel alignment. 
\section{PROPOSED METHOD}\label{sec:method}

\label{sec:pagestyle}

\subsection{Problem Formulation}
Given an unstained dark-field image $\mathbf{x}$, digital staining aims to obtain the stained bright-field image $\mathbf{y}$ with better visibility. In this work, we recast the digital staining task as the sequential light enhancement and image colorization problems which is formulated as
\begin{equation}
    \mathbf{y} = \mathcal{G}(\mathcal{H}(\mathbf{x})) \;,
\end{equation}
where $\mathcal{H}(\cdot)$ and $\mathcal{G}(\cdot)$ denote the light enhancement and colorization modules, respectively, which can sequentially map the dark-field image $\mathbf{x}$ to the corresponding bright-field grayscale image $\mathbf{z}=\mathcal{H}(\mathbf{x})$ and then transfer it into the stained counterpart $\mathbf{y}=\mathcal{G}(\mathbf{z})$.

Figure~\ref{fig:baseline} shows our proposed digital staining pipeline, including a light enhancement module as $\mathcal{H}(\cdot)$, correspondingly, to enhance the input dark-field unstained image (details refer to Section~\ref{sec:tone_mapping}).
After that, a deep generator $\mathcal{G}(\cdot)$ is to map from the unstained to stained domain.
To preserve the structural details, we propose a color-inversion distillation mechanism to ensure data fidelity using the unsupervised strategy (details refer to Section~\ref{sec:distillation}).

\subsection{Histogram Matching for Light Enhancement\label{sec:tone_mapping}}

The cell regions are usually brighter than the background in dark-field unstained images, while they are darker than the background in the bright-field stained images.
Thus, we have the assumption that the cell regions in dark-field unstained images and  bright-field stained images have an approximately reverse illuminance relationship. We propose to match the illuminance distribution of enhanced outputs to that of the bright-field images.

\begin{figure}[t] 
\centering 
\includegraphics[width=0.48\textwidth]{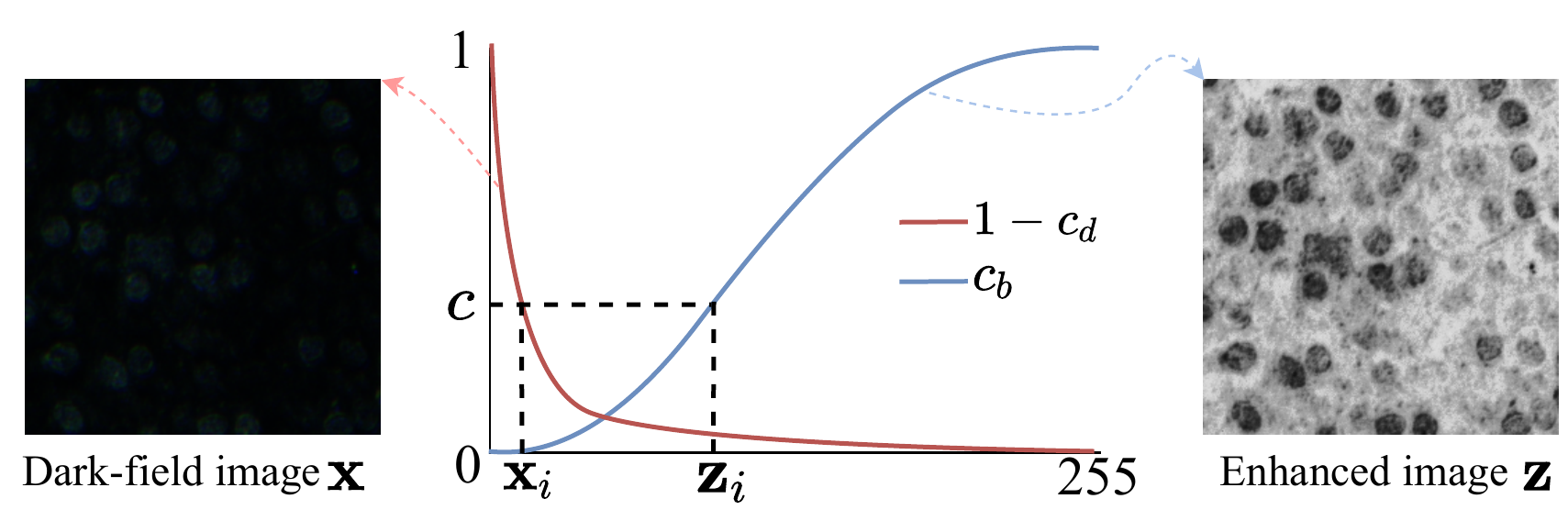} 
\vspace{-4mm}
\caption{Light Enhancement via histogram matching from dark-field to enhanced images. $c$ is the same for both $1-c_d(\mathbf{x}_i)$ and $c_b(\mathbf{z}_i)$. }
\label{fig:res} 
\end{figure}

Histogram matching is a classic method which is widely used in imaging processing tasks such as camera color correction\cite{ding2020multi} and SDR-to-HDR imaging and tone mapping~\cite{bihan2019inverse}. In our task, the mapping from the illuminance distribution of dark-field images to bright-field images is done in an inverse way. 
We use $c_b$ and $c_d$ to indicate the cumulative distribution function of illuminance distribution of bright-field and dark-field images respectively.
For pixel illuminance value $\mathbf{x}_i$ from $\mathbf{x}$ where $i$ represents the pixel coordinates, let $\mathbf{z}_i$ be the grayscale intensity of the corresponding pixel in enhanced image $\mathbf{z}$. Then we have 
$c_{b}(\mathbf{z}_i)=1-c_{d}(\mathbf{x}_i)$, so $\mathbf{z}_i=c_{b}^{-1}(1-c_{d}(\mathbf{x}_i))$.
Based on such an intensity mapping, we can obtain the enhanced cell images as $\mathbf{z}=\mathcal{H}(\mathbf{x})$, leading to much better visibility comparing to the dark-field unstained images as shown in Figure \ref{fig:res}. Thus, the enhanced $\mathbf{z}$ will be used as the input for the following stage.

\begin{figure*}[!t] 
\centering 
\vspace{-10mm}
\includegraphics[width=0.75\textwidth]{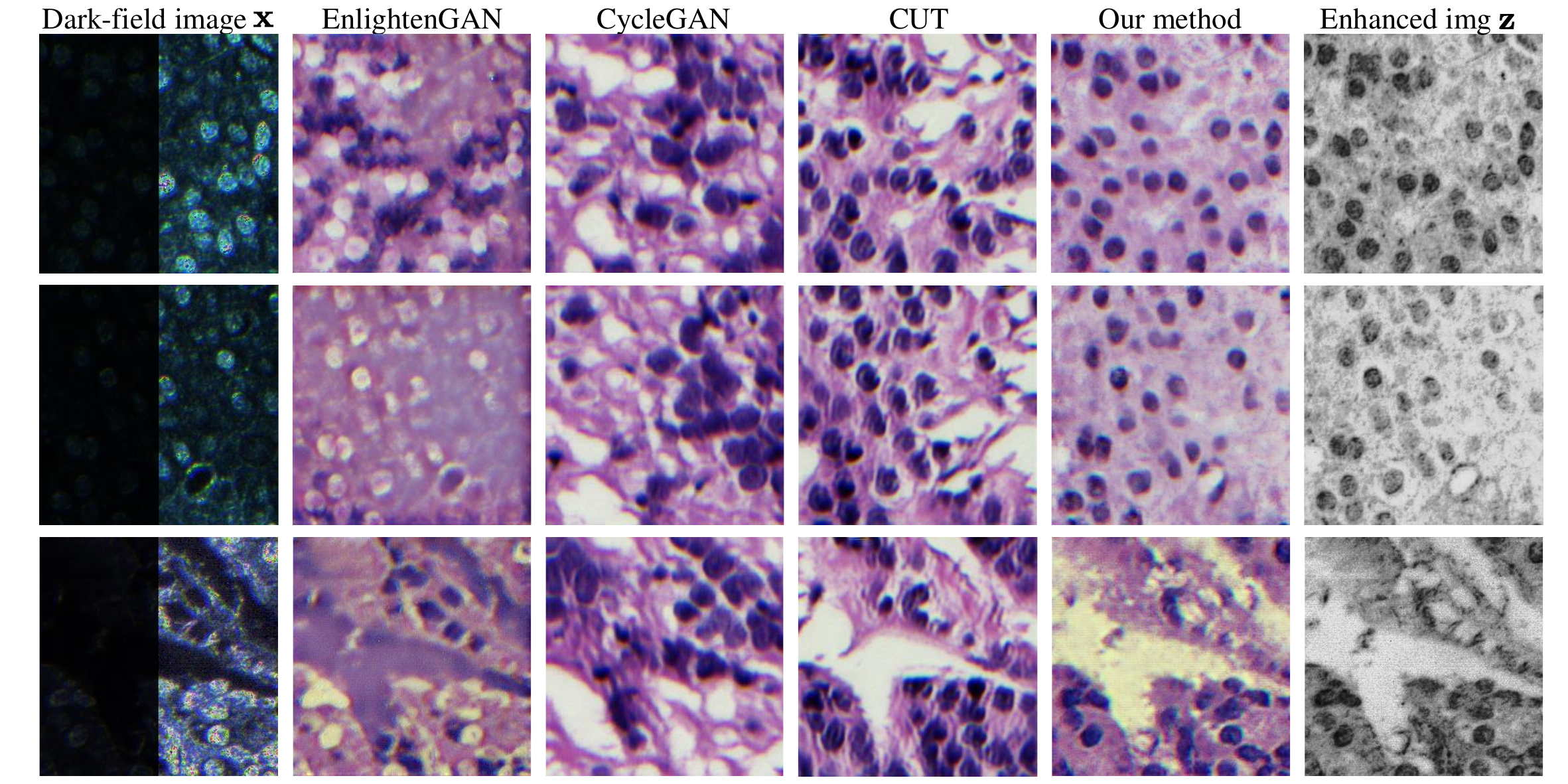} 
\vspace{-3mm}
\caption{
Visual comparison of digital staining results over DSD using our proposed method, as well as other deep unsupervised methods, including
EnlightenGAN\cite{jiang2021enlightengan}, CycleGAN\cite{zhu2017unpaired}, and CUT\cite{park2020contrastive}. For each of dark-field images (leftmost column), the illuminance of its right half part is increased by $\times10$ for better visualization.} 
\label{fig:result_unsupervised} 
\end{figure*}

\subsection{Color-Inversion Distillation for Data Fidelity\label{sec:distillation}}

To preserve image structural information, we propose to employ a pre-trained colorization network $\mathcal{G}_t$ as the teacher model. 
As shown in Figure~\ref{fig:baseline}, we simulate the gray-color training pairs according to inverse the existing stained images into gray (unstained) counterparts.
These synthetic data will be used to pre-train the teacher model $\mathcal{G}_t$ using pixel-wise $\ell_1$ loss.
Given an enhanced cell image $\mathbf{z}$ as the input, we simultaneously feed them to the teacher model $\mathcal{G}_t$ and student model $\mathcal{G}$, achieving the outputs $\mathbf{y}_t$ and $\mathbf{y}$, respectively.
We use a knowledge distillation loss $\mathcal{L}_{kd}$ to minimize the error distance between $\mathbf{y}_t$ and $\mathbf{y}$, which is defined as the following:
\begin{equation}
    \mathcal{L}_{kd} = \|\mathcal{G}(\mathbf{z})-\mathcal{G}_t(\mathbf{z})\|_1 \;.
\end{equation}
Besides, we apply the multi-scale discriminators as the extra supervision to encourage the output to be visually similar to the real stained images, by solving the following problem:
\begin{equation}
    \mathcal{L}_{adv}= \mathop{\sum}_{k=1,2}\mathcal{L}_{GAN}(\mathcal{G},\mathcal{D}_k) \;,
\end{equation}
where $\mathcal{L}_{GAN}$ follows the EnlightenGAN~\cite{jiang2021enlightengan}, $\mathcal{D}_1$ is the global discriminator, $\mathcal{D}_2$ is the local discriminator.
Note that, we only use the distilled student network in the inference stage.
Besides, to further align the cells' position in output images $\mathbf{y}$ with the enhanced images $\mathbf{z}$, we employ the content consistency loss $\mathcal{L}_{con}$ between enhanced images $\mathbf{z}$ and the grayscale image of predicted digital staining image $\mathbf{y}_{gray}$ as follows:
\begin{equation}
  \mathcal{L}_{con} = \|\mathbf{z}-\mathbf{y}_{gray}\|_1 \;.
\end{equation}

By combining the above losses, the hybrid objective function $\mathcal{L}$ that we use to train the student model is formulated as
\begin{equation}
\mathcal{L} = \mathcal{L}_{adv}+\lambda_1\mathcal{L}_{kd}+\lambda_{2}\mathcal{L}_{con} \; ,
\end{equation}
where $\lambda_1$ and $\lambda_2$ are the weighting coefficients to balance the influence of each term.

\section{Experiments}
\subsection{Implementation Details and Setups}

\textbf{Dataset preparation.} 
As previous digital staining works haven't shared their dataset and there's no public datasets with unstained dark-field images and stained bright-field images of same human tissue to the best of our knowledge, we collect a new \textbf{D}igital \textbf{S}taining \textbf{D}ataset, dubbed as \textbf{DSD}, with two unpaired subsets, \ie, unstained dark-field and stained bright-field image subsets, as shown in Figure~\ref{fig:dataset_samples}.
Our training set contains $559$ unstained dark-field images and stained bright-field image pairs with $256\times 256$ resolutions taken from different tissue structures. The teacher model $\mathcal{G}_t$ is trained by bright-field images in the training set.
To better evaluate the performances of our proposed algorithm, we also collect a testing set including $40$ images with the same resolution of the training images.

\begin{table}[!t]
\centering
\footnotesize
\renewcommand{\arraystretch}{0.8}
\adjustbox{width=.9\linewidth}{
    \begin{tabular}{l|cccc}
        \toprule
         Method & FID$\downarrow$ & KID$\downarrow$ &  NIQE$\downarrow$ &LPIPS$\downarrow$ \\\midrule
        EnlightenGAN~\cite{jiang2021enlightengan} & 172.47 & 0.09868& 24.7072 &  {\color{blue}0.46890}\\
         CycleGAN~\cite{zhu2017unpaired} &  {\color{blue}150.03}  & {\color{blue}0.09356} & {\color{blue}8.9616} & 0.65444 \\ 
                  CUT~\cite{park2020contrastive} & 164.70  & 0.09704 & 20.7457&0.59508  \\ 
          \rowcolor{Gray}Ours & {\color{red}\textbf{147.34}}&{\color{red}\textbf{0.08120}} &{\color{red}\textbf{7.1665}} & {\color{red}\textbf{0.33964}}  \\
        \bottomrule
    \end{tabular}
}
    \caption{Quantitative evaluation of the digital staining results over the proposed DSD dataset. }
\label{tab:res}
\end{table}

\vspace{1mm}
\noindent \textbf{Implementation details.} 
We have implemented the proposed model using PyTorch~\cite{paszke2019pytorch}. 
The generator network $\mathcal{G}$ is based on basic blocks from ResNet~\cite{he2016deep}. $\mathcal{G}$ consists of one convolution, two 2-stride convolutional layers , nine residual blocks and two fractionally-strided convolutions with stride $\frac{1}{2}$ and one convolution inspired by~\cite{zhu2017unpaired}, which has impressive effects in many image-to-image translation tasks.  For teacher colorization network $\mathcal{G}_t$, we use a U-Net based network following~\cite{zhang2017real}. The teacher network is trained with 559 bright-field images in the training set.
 We adopt Adam optimizer~\cite{kingma2014adam} for loss minimization, with initial learning rate set to 0.0001 for 200 epochs, followed by another 100 epochs with learning rate linearly decayed to 0.  For weight coefficients, we set $\lambda_1=\lambda_2=10$.
 
\vspace{1mm}
\noindent \textbf{Evaluation metrics.} 
We opt for perceptual metrics including NIQE~\cite{mittal2012making}, FID~\cite{heusel2017gans}, and KID~\cite{binkowskidemystifying} for evaluation.
Due to the domain gap between cell images and natural images, we employ the stained bright-field image set as the reference.
Besides, we also adopt LPIPS~\cite{zhang2018unreasonable} to evaluate the performance on preserving the content information from the input image. Specifically, we calculate the LPIPS between the grayscale image of predicted bright-field image $\mathbf{y}_{gray}$ and the corresponding enhanced image $\mathbf{z}$.
For all these metrics, a lower value indicates better visual quality. 

\subsection{Comparison With Existing Methods}
As we are the first work of unsupervised method on digital staining, no existing digital staining methods can compare to.
We select several recent unsupervised image-to-image translation methods, \ie, EnlightenGAN~\cite{jiang2021enlightengan}, CycleGAN~\cite{zhu2017unpaired}, and CUT~\cite{park2020contrastive}, as the competing methods.
For a fair comparision, we re-trained all competing methods on the training set of our DSD dataset.
Table~\ref{tab:res} shows the quantitative results, in which the proposed method outperforms all competing methods for both image quality and data diversity.
The visual examples have been demonstrated in Figure~\ref{fig:result_unsupervised}. 
EnlightenGAN and CycleGAN fail to distinguish cells and background part, thus predict cavity where supposed to be cells. CUT predicts cells with rough edges and fake cavity areas. 
Our method can generate the stained bright-field image with complete structural information, like position and shape of cells and cavities. 

\subsection{Ablation Study} 
\begin{figure}[!t] 
\centering 
\includegraphics[width=1.\linewidth]{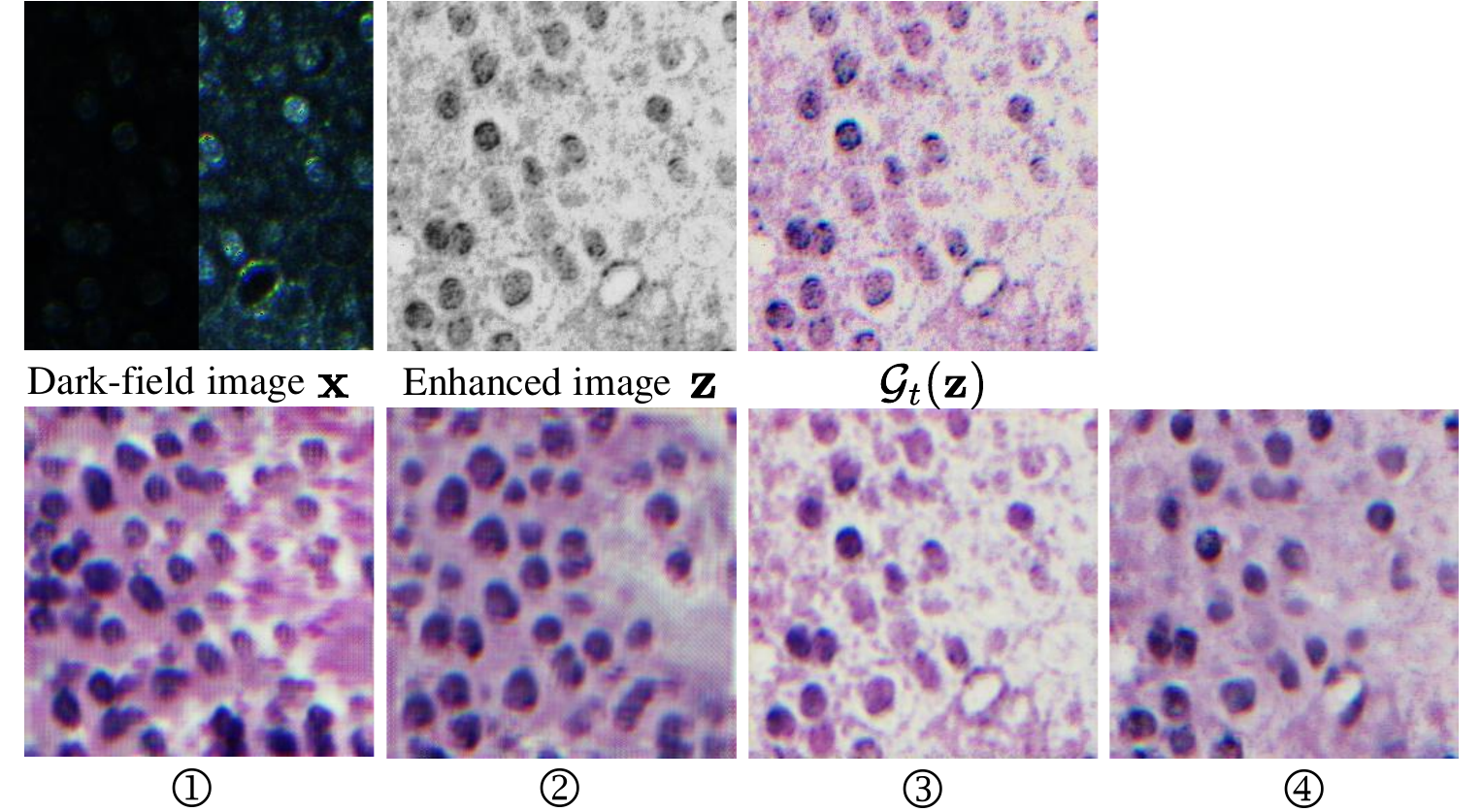} 
\vspace{-6mm}
\caption{
Visual examples of ablation study, including input dark-field image $\mathbf{x}$ (left) and its corresponding $\times 10$ version (right) in the top-left subfigure, the enhanced image $\mathbf{z}$, and the results from teacher network $\mathcal{G}_t(\mathbf{z})$. Results of four ablation experiments correspond to the numerical order in Table~\ref{tab:ablation}.}
\label{fig:result} 
\end{figure}

\noindent\textbf{The effect of different loss functions.}
We thoroughly investigate the impact of each loss function applied in the training stage.
Table~\ref{tab:ablation} shows the evaluation results and Figure~\ref{fig:res} demonstrates the visual examples with different loss functions.
We first remove the knowledge distillation loss $\mathcal{L}_{kd}$ and replace the content consistency loss $\mathcal{L}_{con}$ with the VGG-based perceptual loss $\mathcal{L'}_{con}$, denoted as \ding{172} in Table~\ref{tab:ablation}. Here the applied perceptual loss can be written as
\begin{equation}
  \mathcal{L'}_{con} = \|\phi_n(\mathbf{z})-\phi_n(\mathbf{y}_{gray})\|_1\;,\end{equation}
where $\phi_n$ denotes the n-th layer map of the VGG-16 model pretrained on ImageNet.
In our experiment, we choose the first convolutional layer after the fifth pooling layer of VGG-16 and the weight assigned to $\mathcal{L'}_{con}$ equals to 1 following the settings from ~\cite{jiang2021enlightengan}.
We find that the position of the cells has almost all shifted with only perceptual loss as shown in Figure~\ref{fig:res}.
Then we verify the effectiveness of the knowledge distillation loss $\mathcal{L}_{kd}$, in which we remove the $\mathcal{L}_{kd}$ and preserve the content consistency loss $\mathcal{L}_{con}$, denoted as \ding{173} in the Figure~\ref{fig:res}.
The shape and size of cells still cannot be consistent with the original dark-field image if removing the $\mathcal{L}_{kd}$, comparing \ding{173} and \ding{175} in Figure~\ref{fig:res}.
Besides, according to the quantitative results in Table~\ref{tab:ablation}, the stained image quality and diversity would be degraded.

\vspace{1mm}
\noindent\textbf{The effect of different student model.}
To justify our choice of student model, we compare the performance of our method with the student model replaced with the generator followed by ~\cite{jiang2021enlightengan}.
The corresponding result is shown in \ding{174} of Figure~\ref{fig:result} and Table~\ref{tab:ablation}. 
We can see the choice of generator network improves the visual quality in background area in some images especially. 

\begin{table}[!t]
\centering
\footnotesize
    \setlength{\tabcolsep}{0.4em}
\renewcommand{\arraystretch}{0.8}
\adjustbox{width=1.\linewidth}{
    \begin{tabular}{c|ccc|c|cccc}
        \toprule
       &$\mathcal{L}_{con}$ & $\mathcal{L'}_{con}$ & $\mathcal{L}_{kd}$ & Type of $\mathcal{G}$ & FID$\downarrow$ & KID$\downarrow$ &  NIQE$\downarrow$ &LPIPS$\downarrow$  \\ \midrule
    \ding{172} & & \Checkmark& & ResNet~\cite{he2016deep} & 172.80 &  0.13479 & 9.7733 & 0.54746\\
       \ding{173} &\Checkmark & 
       & & ResNet~\cite{he2016deep} & 181.84 & 0.14736  & {\color{blue}{9.3443}} & 0.50864\\
        \ding{174} &\Checkmark &  &\Checkmark & EG~\cite{jiang2021enlightengan} & {\color{blue}{159.19}} & {\color{blue}{0.08621}} & 10.5326 & {\color{red}{\textbf{0.31524}}} \\
      \rowcolor{Gray} \ding{175} & \Checkmark &  & \Checkmark & ResNet~\cite{he2016deep}&  {\color{red}\textbf{147.34}} &{\color{red}\textbf{0.08120}} &{\color{red}\textbf{7.1665}} & {\color{blue}{0.33964}}\\

        \bottomrule
    \end{tabular}
}
    \caption{Quantitative comparison for the ablation study. The results of the model  \ding{172} with only the VGG-based perceptual loss $\mathcal{L'}_{con}$, \ding{173} with only the content consistency loss,  \ding{174} replacing the student model using EG (EnlightenGAN~\cite{jiang2021enlightengan}), and \ding{175} the complete model. Note that all variant models include the adversarial loss $\mathcal{L}_{adv}$.}
\label{tab:ablation}
\end{table}

\section{Conclusion}
In this paper we propose a novel unsupervised digital staining method learning from unstained dark-field images to H\&E stained bright-field images. We use knowledge distillation loss and content consistency loss to preserve the structural information consistency, and generative adversarial network architecture to encourage the output visually similar with real stained images. With unsupervised learning framework, our method can be trained with unpaired data. Experimental results also demonstrate that our method achieve the superior performance compared to all competing methods and variants, 
with the style of H\&E stained images while keeping the structural information. We propose and will release a new dataset of unpaired
real dark-field unstained and bright-field stained cell images.

\bibliographystyle{IEEEbib}
\bibliography{strings,refs}
\
\end{document}